\documentclass
[superscriptaddress,secnumarabic,amssymb,amsmath,nobibnotes,aps,prd,showkeys,showpacs,nofootinbib,onecolumn]{revtex4}%
\usepackage{graphicx}
\usepackage{epsf}
\usepackage{bm}
\usepackage{amsmath}
\usepackage{amsfonts}
\usepackage{amssymb}%
\providecommand{\U}[1]{\protect\rule{.1in}{.1in}}
\newcommand{\be}{\begin{equation}}
\newcommand{\ee}{\end{equation}}

\newcommand{\mincir}{\raise
-3.truept\hbox{\rlap{\hbox{$\sim$}}\raise4.truept\hbox{$<$}\ }}
\newcommand{\magcir}{\raise
-3.truept\hbox{\rlap{\hbox{$\sim$}}\raise4.truept\hbox{$>$}\ }}

\begin{document}
\title{Cosmological singularities and analytical solutions in varying vacuum cosmologies}
\author{Spyros Basilakos}
\email{svasil@academyofathens.gr}
\affiliation{Academy of Athens, Research Center for Astronomy and Applied Mathematics,
Soranou Efesiou 4, 11527, Athens, Greece}
\author{Andronikos Paliathanasis}
\email{anpaliat@phys.uoa.gr}
\affiliation{Instituto de Ciencias F\'{\i}sicas y Matem\'{a}ticas, Universidad Austral de
Chile, Valdivia, Chile}
\affiliation{Institute of Systems Science, Durban University of Technology, PO Box 1334,
Durban 4000, Republic of South Africa}
\author{John D. Barrow}
\email{jdb34@hermes.cam.ac.uk}
\affiliation{DAMTP, Centre for Mathematical Sciences, University of Cambridge, Wilberforce
Rd., Cambridge CB3 0WA, UK}
\author{G. Papagiannopoulos}
\email{yiannis.papayiannopoulos@gmail.com}
\affiliation{Faculty of Physics, Department of Astronomy-Astrophysics-Mechanics University
of Athens, Panepistemiopolis, Athens 157 83, Greece}
\keywords{Cosmology; $\Lambda$-varying; Integrability; Singularity}
\pacs{98.80.-k, 95.35.+d, 95.36.+x}

\begin{abstract}
We investigate the dynamical features of a large family of running vacuum
cosmologies for which $\Lambda$ evolves as a polynomial in the Hubble
parameter. Specifically, using the critical point analysis we study the
existence and the stability of singular solutions which describe de-Sitter,
radiation and matter dominated eras. We find several classes of $\Lambda(H)$
cosmologies for which new analytical solutions are given in terms of Laurent
expansions. Finally, we show that the Milne universe and the $R_{h}=ct$ model
can be seen as perturbations around a specific $\Lambda(H)$ model, but this
model is unstable.

\end{abstract}
\maketitle
\date{\today}

\
%\author{Andronikos Paliathanasis}

%\author{Spyros Basilakos}

%\author{G. Papagiannopoulos}

%\author{John D. Barrow}

\section{Introduction}

Over the last two decades, most studies in cosmology strongly indicate that
the universe is spatially flat and it consists of $\sim4\%$ baryonic matter,
$\sim26\%$ dark matter and $\sim70\%$ of dark energy (hereafter DE), thought
to be driving the phenomenon of cosmic acceleration
\cite{dataacc1,dataacc2,data1,data2,data3,data4}.
%A straightforward and easy
%way to explain this expansion is to consider an additional fluid, that we have
%come to call Dark Energy (DE) and according to recent data consists 70\% of
%the total energy of the Universe.
Although there is a general consensus regarding the main properties of DE,
namely it has a negative pressure, the origin of this unexpected component of
the universe has yet to be identified. This has given rise to a plethora of
alternative cosmological scenarios which mainly generalize the nominal
Einstein-Hilbert action of general relativity using either a new field in
Nature \cite{hor1,hor2,hor3,hor4}, or a non-standard gravity theory that
increases the number of degrees of freedom
\cite{clifton,mod1,mod2,mod3,mod4,mod5}.
%The novelty in the modified theories
%of gravity is that they provide possible scenarios that would explain the
%accelerating expansion of the universe without the need to introduce a dark
%energy fluid, that is, the acceleration of the universe is one of geometric origin.

The introduction of a cosmological constant, $\Lambda$, is probably the
simplest modification of the Einstein-Hilbert action which can be considered.
In the framework of the so called $\Lambda$CDM model, the cosmological
constant coexists with cold dark matter (CDM) and ordinary baryonic matter
(see \cite{Pee2003} for review). Although the $\Lambda$CDM model fits
accurately, the current cosmological data suffers from two problems
\cite{conpr1,conpr2,conpr3,conpr4}. The first is the 'old' cosmological
problem, namely the expected (Planck natural unit) vacuum energy density is
$\sim120$ orders of magnitude larger than the presently observed value of
$\Lambda$. The second problem is the coincidence problem: that the density of
dark energy is so similar to the matter density today (the two were equal when
the universe had expanded to about 75\% of it present expansion scale).

An alternative approach to resolving these two problems is to consider the so
called $\Lambda\left(  t\right)  $CDM models, wherein $\Lambda$ is allowed to
vary with cosmic time (see \cite{Bas2009a,Bass1,perico} and references
therein). This class of models
\cite{Ozer,bertolami,chenwu,lim00,lima6,lima7,many1,many2,many3,aldrovandi,Schutz,Schutz2,Lima1,Lima2,Lima4,Lima5,SalimWaga,Waga,span1}
is based on a dynamical vacuum energy density that evolves as a power series
of the Hubble rate (for review see \cite{ShapiroSola}, \cite{Sool14}). Powered
by a decaying vacuum energy density, the spacetime can emerges from a pure
non-singular initial de Sitter vacuum stage, "gracefully" exit from inflation
to a radiation era followed by dark-matter and vacuum-dominated phases, before
finally, evolving to a late-time de Sitter phase \cite{bas22,Bass1,perico}.
Recently, Sola et al. \cite{sol16} tested the performance of the running
vacuum models against the latest cosmological data and they found that the
$\Lambda(H)$ models are favored than the usual $\Lambda$CDM model at
$\sim4\sigma$ statistical level (see also \cite{zhao17}).
%Of course, these
%models have more parameters than $\Lambda$CDM and must therefore be able to
%fit the data better.
These developments have led to growing interest in $\Lambda(H)$ cosmological models.

There was a great effort to explore the $\Lambda(H)$ models both analytically
and observationally but a dynamical analysis based on critical points is still
missing. The purpose of our work is to bridge this gap, by determining the de
Sitter phases of a general family of $\Lambda(H)$ models to search for
singular solutions of the form $a\left(  t\right)  \propto t^{p}$ which secure
the existence of radiation ($p=1/3$) and matter ($p=2/3$) dominated eras,
respectively. We will investigate the stability of the critical points in
order to understand the dynamical and cosmological properties of the
$\Lambda(H)$ models. Here, the main mathematical tool that we use is that of
the singularity analysis of differential equations, and more specifically we
apply the ARS (Ablowitz, Ramani and Segur) algorithm \cite{Abl1,Abl2,Abl3}.
This algorithm provides a method to construct the analytical solution of a
differential equation which is expressed as a Laurent expansion around the
singular leading-order term (for some applications on gravitational theories
see \cite{sin1,sin2,sin3,sin4,sin5,sin6} and references therein). Information
regarding the stability of the trajectories close to the singularity can be
extracted directly from the ARS algorithm.
%the latter is important in order to
%understand the evolution of the universe as we exit from the matter dominated
%era.

The structure of the manuscript is as follows. In Section \ref{sec2} we
briefly introduce the concept of the running $\Lambda(H)$ cosmologies. In
sections 3 and 4 we present the main results of our work, namely we study the
critical points and their stability as well as we provide the corresponding
analytical solutions. Finally, in section \ref{con} we draw our conclusions.

\section{$\Lambda$-varying Cosmology}

\label{sec2}

%Nonetheless not all
%possible functional dependences on $H$ are allowed. In order to obtain a
%definite decaying cosmology we need to find a viable expression in terms of
%the Hubble rate. Over the last years there have been several different
%versions of the $\Lambda(H)$ models, some of which are: (a) the
%Renormalization Group (RG)\ based vacuum model of Quantum Field theory
%\ ($\Lambda_{RG}=n_{0}+n_{2}H^{2}$), (b) the vacuum model where ($\Lambda
%_{H}\varpropto H^{2}$), (c) Power series expansions models \ ($\Lambda
%_{ps1}=n_{1}H+n_{2}H^{2},\Lambda_{ps2}=n_{1}H$).

%These models can be tested and constrained by a variety of observations such
%as Baryonic Acoustic Oscillations, the Cosmic Microwave background shift
%parameter as well as SNIa distance moduli. While the aforementioned models are
%able to withstand these tests \cite{bassola} those mostly favored are
%the\textbf{ }$\Lambda$CDM and the $\Lambda_{RG}.$In this work we assume
%time-varying vacuum energy density of the more inclusive form $\Lambda
%=\Lambda\left(  H\right)  $\ mixed with perfect fluids. We will also treat the
%vacuum density as that of a perfect fluid.

In this section we briefly present the main points of the running vacuum
cosmology. If we model the expanding universe as a perfect fluid with density
$\rho$, and corresponding pressure\footnote{In the case of $w=0$ we have
non-relativistic matter, while for $w=1/3$ we have relativistic matter
(radiation).} $p=w\rho$, then the energy-momentum tensor is given by ${T}%
_{\mu\nu}=-p\,g_{\mu\nu}+(\rho+p)U_{\mu}U_{\nu}$. In this context, the term
$\Lambda\,g_{\mu\nu}$ can be absorbed by the total energy momentum tensor
$\tilde{T}_{\mu\nu}\equiv T_{\mu\nu}+g_{\mu\nu}\rho_{\Lambda}$, where
$\rho_{\Lambda}=\Lambda/(8\pi G)$ is the vacuum energy density which is
related to $\Lambda$, while the corresponding equation of state is
$p_{\Lambda}=-\rho_{\Lambda}$. Combining the above expressions we arrive at
\begin{equation}
\tilde{T}_{\mu\nu}=(\rho_{\Lambda}-p)\,g_{\mu\nu}+(\rho_{m}+p)U_{\mu}U_{\nu
}\,, \label{Tmunuideal}%
\end{equation}
and thus the Einstein's field equations become
\begin{equation}
R_{\mu\nu}-\frac{1}{2}g_{\mu\nu}R=8\pi G\ \tilde{T}_{\mu\nu}\,. \label{EE}%
\end{equation}
For the spatially flat FLRW metric the field equations boil down to those of
Friedmann equations, namely
\begin{equation}
3H^{2}\left(  t\right)  =\Lambda\left(  t\right)  +\rho\left(  t\right)  ,
\label{fe.01}%
\end{equation}%
\begin{equation}
-2\dot{H}\left(  t\right)  -3H^{2}\left(  t\right)  =-\Lambda\left(  t\right)
+p\left(  t\right)  , \label{fe.02}%
\end{equation}
where for simplicity we have set $8\pi G\equiv1$. Although, a constant
$\Lambda$ term is the simplest possibility, it is interesting to mention that
the Cosmological Principle embodied in the FLRW spacetime allows $\Lambda$ to
be a function of the cosmic time, or of any collection of homogeneous and
isotropic dynamical variables, i.e. $\Lambda=\Lambda(\chi(t))$. Moreover,
considering $G=const.$ (for other scenarios see \cite{Sool14}) the Bianchi
identity that insures the covariance of the formulation, implies an energy
exchange between $\rho_{\Lambda}$ (or $\Lambda$) and $\rho$
\begin{equation}
\dot{\rho}+\dot{\Lambda}+3H\left(  \rho+p\right)  =0. \label{fe.03}%
\end{equation}
%The above equations (\ref{fe.01})-(\ref{fe.03}) are not independent, since
%from Eq.(\ref{fe.02}) and Eq.(\ref{fe.01}) we can extract Eq.(\ref{fe.03}). We
%can see there exists a simple exact scaling solution in which all the terms in
%(\ref{fe.01}) and $p=(\gamma-1)\rho$, with $\gamma$ constant, are proportional
%to $t^{-2}$. This gives an exact solution of (\ref{fe.01})-(\ref{fe.03}):%

%\begin{equation}
%H(t)=\frac{h}{t}~,~~\rho(t)=\frac{2h}{\gamma t^{2}}~,~~\Lambda(t)=\frac
%{h(3h\gamma-2)}{\gamma t^{2}}. \label{sol1}%
%\end{equation}
%Hence, the ratio of the dark energy density to the other matter densities are
%always constant, and given by%

%\[
%\frac{\Lambda}{\rho}=\frac{3h\gamma-2}{2}.
%\]
%Observationally, this ratio is about $\frac{\Lambda}{\rho}\simeq7/3$ which, in
%the dust ($\gamma=1$) era, would require $h=20/9.$This solution contains only
%one parameter ($h$) after the equation of state ($\gamma$) is fixed and it has
%the interesting 'everlasting lambda' property that the cosmological constant
%is always important because it is always of similar order to that of $\rho$.
%Another scenario of this sort has been proposed but turned out to make
%predictions about spatial fluctuations in conflict with current observations
%\cite{jb1}. A different approach that does not have this problem is introduced
%in refs. \cite{jb2, jb3}.\emph{ }

In order to find more general solutions for this scenario, ones in which there
is evolution of the form of the expansion rate over time, we need to explore a
more general functional form for $\Lambda(H)$. The case of viable running
vacuum $\Lambda$ can be placed in the general framework of quantum field
theory in curved spacetime \cite{JSola,ShapiroSola,StarobinskyPhysLet}.
Specifically, in ref. \cite{perico} the following expression (for recent
review see \cite{gomez}) was proposed for the functional form of
$\Lambda(H),$
\begin{equation}
\Lambda\left(  H\right)  =c_{0}+c_{1}H^{2}+c_{2}H^{2k}, \label{fe.0333}%
\end{equation}
where $k>0$. Notice, that this formula is well motivated in the general
context of the re-normalization quantum field theory (QFT) in curved spacetime
and thus only even powers in $H$ are allowed by the general covariance of the
theory (see \cite{ShapiroSola}, \cite{Sool14}). In this family of running
vacuum models the spacetime emerges from a pure non-singular de Sitter vacuum
stage, allowing a graceful exit from inflation to a radiation-dominated phase,
followed by dark matter and vacuum regimes, followed by evolution to a
late-time de Sitter phase \cite{gomez,Bass1}. From the observational
viewpoint, the current vacuum model agrees with the latest expansion data and
it predicts a growth rate of clustering which is in agreement with the
observations (for more details see \cite{gomez,gomez1,sol16} and references therein).

Inspired by ref. \cite{perico} and based only on phenomenological arguments we
introduce in our analysis the following form of the running vacuum
\begin{equation}
\Lambda\left(  H\right)  =c_{0}+c_{1}H^{2}+c_{2}H^{-n}, \label{fe.033}%
\end{equation}
where the leading term $c_{0}$ describes in good approximation the current
universe and the other terms introduce a dynamical evolution which can be
important under special conditions. Unlike Eq.(\ref{fe.0333}), here the
exponent $n$ is allowed to take positive and negative values as well as it can
be either even or odd. From the phenomenological point of view, the
parametrization (\ref{fe.033}) describes a large family of dynamical models of
the vacuum energy (see also \cite{gomez}). As expected, for the special case
where $n=-2k$ the current form of $\Lambda(H)$ reduces to that of
Eq.(\ref{fe.0333}).

Furthermore, for a general quadratic vacuum model, with $\Lambda\left(
H\right)  =c_{1}H^{2}+F\left(  H\right)  ,$ we find
\begin{equation}
2\dot{H}+3\left[  \left(  1+w\right)  (1-\frac{c_{1}}{3})\right]
H^{2}-\left(  1+w\right)  F\left(  H\right)  =0, \label{fe.master}%
\end{equation}
where $F\left(  H\right)  $ is an arbitrary function. Introducing the
following parametrizations,
\begin{equation}
3\left[  \left(  1+w\right)  (1-\frac{c_{1}}{3})\right]  =3\left(
1+w_{1}\right)  ,\;\;\;\;\;F(H)=\frac{\left(  1+w_{1}\right)  }{\left(
1+w\right)  }\bar{F}\left(  H\right)  , \label{fe.master00}%
\end{equation}
Eq.(\ref{fe.master}) then becomes
\begin{equation}
2\dot{H}+3\left(  1+w_{1}\right)  H^{2}-\left(  1+w_{1}\right)  \bar{F}\left(
H\right)  =0. \label{fe.master1}%
\end{equation}
This equation is just the original master equation (\ref{fe.master}) with a
different value for $w$, which means that the quadratic term $H^{2}$ of
Eq.(\ref{fe.033}) can be easily absorbed into the Friedmann equation
(\ref{fe.master}). In this context, if we decide to use Eq.(\ref{fe.master1})
in order to study the current class of decaying vacuum models, then
$\Lambda(H)$ takes the form
%so that the effects of
%the
%quadratic term $H^{2}$ in Eq.(\ref{fe.033})
%are canceled. Therefore, without loss of generality we
%consider $c_{1}\rightarrow0$, \ where the latter $\Lambda-$varying model takes
%the following form%
\begin{equation}
\Lambda\left(  H\right)  =\lambda_{0}+\lambda_{2}H^{-n},\;\;\;n\neq-2,
\label{fr.03a}%
\end{equation}
where the new constants $\lambda_{0}$ and $~\lambda_{2}$ are related with
those of (\ref{fe.033}) by $c_{0}=\frac{3-c_{1}}{3}\lambda_{0}$ and
$c_{2}=\frac{3-c_{1}}{3}\lambda_{2}$. At this point, it is important to
mention that the equation of state parameter for the matter source has been
changed as given by (\ref{fe.master00}).

In order to understand the possible evolutionary tracks of the cosmic fluid,
we consider the following two cases:

\textbf{Case A}: Assume that the matter source is that of an ideal fluid for
which the equation of state parameter $w_{1}$ is constant. Hence, from
Eq.(\ref{fe.master1}), we deduce that
\begin{equation}
2\dot{H}+3\left(  1+w_{1}\right)  H^{2}-\left(  1+w_{1}\right)  \left(
\lambda_{0}+\lambda_{2}H^{-n}\right)  =0, \label{fr.03b}%
\end{equation}
or, with the aid of $H={\dot{a}}/a,$ we find
\begin{equation}
2a\ddot{a}+\left(  1+3w_{1}\right)  \dot{a}^{2}-\left(  1+w_{1}\right)
a^{2}\left[  \lambda_{0}+\lambda_{2}\left(  \dot{a}a^{-1}\right)
^{-n}\right]  =0. \label{fr.03c}%
\end{equation}

\begin{figure}[ptb]
\includegraphics[height=6.5cm]{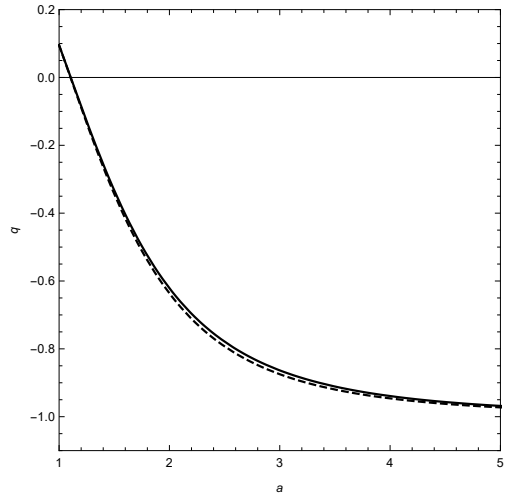}\centering\caption{Qualitative
evolution of the deceleration parameter $q\left(  a\right)  $, for the vacuum
model (\ref{fr.03c}) for $w_{1}=0$. Solid line is for $n=3$ and $\lambda
_{2}=1$, while dashed line is for $n=1$ and $\lambda_{2}=\lambda_{2}^{B}$ with
$\lambda_{2}^{B}\gg1$. The behaviour of the deceleration parameter is similar
for all the positive values of $\lambda_{2}$ and $n.$\ For the initial
conditions of the numerical simulation we selected $a\left(  t_{0}\right)
=1$, $\dot{a}\left(  t_{0}\right)  =70$ and $\Omega_{m0}\simeq0.3.$}%
\label{qqaa}%
\end{figure}

At this point we would like to discuss the following issue: \textit{for the
vacuum model with $n>0$ is it possible to have a super-accelerated expansion
of the universe in the far future?} In order to answer this question we need
to compute the deceleration parameter $q=-(1+\frac{\dot{H}}{H^{2}})$ by
numerically integrating Eq.(\ref{fr.03c}) for non-relativistic matter
$w_{1}=0$. Concerning the initial conditions we use the present values,
mainely, $a\left(  t_{0}\right)  =1$ and ${\dot{a}}\left(  t_{0}\right)  =70$,
$\mathbf{\lambda_{0}}$ $=3\Omega_{m0}H_{0}^{2}-\lambda_{2}\mathbf{H}_{0}%
^{-n},~$with $\Omega_{m0}\simeq0.3$. We find that in the far future the
deceleration parameter tends to that of de-Sitter cosmology ($q\approx-1$) and
thus we exclude the possibility of a super-accelerated expansion of the
universe. In Figure \ref{qqaa} we plot the evolution of the deceleration
parameter for various $\lambda_{2}$ and $n$ values. \textbf{Practically,
$H^{-n}$ with $n>0$ does not really affect the cosmic expansion at late times.
Moreover, one can easily check that in this case the term $H^{-n}$ plays no
role in the early inflation. Under the above conditions in order to provide a
viable running vacuum model one may introduce other additional terms, namely
${\dot H}^{n}$ in the form of $\Lambda(H)$ (see \cite{gomez} and references
therein).}

%In order to study if there is possible the current vacuum model to
%provide a future super-accelerated expansion of the universe we perform a
%numerical simulation of the latter equation in the limit whre }$H^{-n}%
%$\textbf{ dominates over the term }$\lambda_{0}$\textbf{, while for the matter
%source we assume that~}$w_{1}=0$\textbf{. The evalution of the deceleration
%parameter }$q\left(  a\right)  =-(1+\frac{\dot{H}}{H^{2}})$\textbf{ is given
%in Figure \ref{qqaa} for positive and negative values of the parameters
%}$\lambda_{2}$\textbf{ and }$n$\textbf{. We observe that for negative values
%of }$\lambda_{2},$\textbf{ }$q\left(  a\right)  \rightarrow-1$\textbf{, which
%means that the solution behaves like a de Sitter universe, as we shall see in
%the latter section. On the other hand for }$\lambda_{2}>0$\textbf{, the
%solution collapse.

\textbf{Case B}: Here we assume that the cosmic fluid contains two ideal
fluids (nominally case matter and radiation), hence
\begin{equation}
\rho\left(  t\right)  =\rho_{1}\left(  t\right)  +\rho_{2}\left(  t\right)  ,
\label{fe.04}%
\end{equation}
and%
\begin{equation}
p\left(  t\right)  =p_{1}\left(  t\right)  +p_{2}\left(  t\right)  ,
\label{fe.05}%
\end{equation}
where $p_{1}=w_{1}\rho_{1}$ and $p_{2}=w_{2}\rho_{2}$. Moreover, assume the
second fluid does not interact either with the first fluid or with
$\Lambda(H)$.
%, namely $w_{2}$ is
%minimally coupled.
Hence, the conservation law for this component implies ${\dot{\rho}}%
_{2}+3(1+w_{2})H\rho_{2}=0,$ a solution of which is
\begin{equation}
\rho_{2}\left(  t\right)  =\rho_{20}a^{-3\left(  1+w_{2}\right)  },
\label{fe.06}%
\end{equation}
where we have set $w_{2}=$ const. Therefore, it is easy to show that we need
to introduce the evolution of the second fluid into Eq.(\ref{fr.03b}), namely
\begin{equation}
2\dot{H}+3\left(  1+w_{1}\right)  H^{2}-\left(  1+w_{1}\right)  \left(
\lambda_{0}+\lambda_{2}H^{-n}\right)  +w_{2}\rho_{20}a^{-3\left(
1+w_{2}\right)  }=0, \label{fe.07}%
\end{equation}
or
\begin{equation}
2a\ddot{a}+\left(  1+3w_{1}\right)  \dot{a}^{2}-\left(  1+w_{1}\right)
a^{2}\left[  \lambda_{0}+\lambda_{2}\left(  \dot{a}a^{-1}\right)
^{-n}\right]  +w_{2}\rho_{20}a^{-(1+3w_{2})}=0. \label{fe.08}%
\end{equation}
Notice, that the pair $(w_{1},w_{2})$ lies in the region $\left(  -1,1\right)
$.

In the rest of the paper we proceed to determine analytical solutions of the
Friedmann equations (\ref{fr.03c}) and (\ref{fe.08}). Since these equations
have a nonlinear nature we apply the algorithmic method of singularity
analysis due to Ablowitz, Ramani and Segur (ARS) \cite{Abl1,Abl2,Abl3}. This
method provides information about the existence and stability of singular
cosmological solutions in different eras. In the limit where the constant
$\lambda_{0}$ vanishes with $n<0$, the closed-form solution of (\ref{fr.03b})
was found by Perico et al. \cite{perico} (see also \cite{Carn08},
\cite{Bas09a}). However, the existence of singular solutions
%and the perturbations
close to the singularity has not yet been studied. As we shall see below,
%our analysis is valid and in the $\Lambda
%_{0}\rightarrow0$ case. \ However from the expression of the analytical
%solution we shall see that
the constant term, $\lambda_{0},$ modifies the solution close to the
singularity and so affects the entire cosmic history.

\section{De Sitter phases}

\label{sec3}

Consider the scenario with the two ideal fluids, $\rho_{1}\left(  t\right)  $
and $\rho_{2}\left(  t\right)  $. We choose the dimensionless variables
$\Omega_{1}\left(  t\right)  =\frac{\rho_{1}\left(  t\right)  }{3H^{2}%
},~\Omega_{2}\left(  t\right)  =\Omega_{2}\left(  t\right)  =\frac{\rho
_{2}\left(  t\right)  }{3H^{2}}$, and after some algebra the field equations
reduce to the following system of first-order equations:%

\begin{equation}
2\dot{H}=\left(  1+w_{2}\right)  \left(  \lambda_{0}+\lambda_{2}H^{-n}\right)
-3H^{2}\left(  1+w_{2}+\left(  w_{1}-w_{2}\right)  \Omega_{1}\right)  ,
\label{fe.32}%
\end{equation}%
\begin{align}
6\dot{\Omega}_{1}  &  =-18\left(  w_{1}-w_{2}\right)  \left(  1-\Omega
_{1}\right)  H\Omega_{1}-6\left(  1+w_{2}\right)  \lambda_{0}\Omega_{1}%
H^{-1}-n\left(  1+w_{2}\right)  \left(  \lambda_{2}\right)  ^{2}%
H^{2n-3}+\nonumber\\
&  ~\ ~-H^{n-3}\left(  n(1+w_{2})\lambda_{0}\lambda_{2}\right)  -\ 3\lambda
_{2}H^{n-1}(-n(1+w_{2})+(2-nw_{1}+(2+n)w_{2})\Omega_{1}), \label{fe.33}%
\end{align}
while the constraint Eq.(\ref{fe.01}) becomes
\begin{equation}
3H^{2}\left(  1-\Omega_{1}-\Omega_{2}\right)  -\lambda_{0}-\lambda_{2}%
H^{-n}=0. \label{fe.34}%
\end{equation}

In order to have a de Sitter phase the corresponding critical point $P=\left(
H_{P,}\Omega_{P}\right)  $ of the dynamical system (\ref{fe.32})-(\ref{fe.33})
needs to obey ${\dot{H}}(P)=0$. This condition implies
\[
H_{P}^{2}\Omega_{1}=0,\;\;\;\;3H_{P}^{2}-\lambda_{0}-H_{P}^{-n}\lambda_{2}=0.
\]
Therefore, if we consider that $H_{P}\neq0$, which is always true for
$\lambda_{0}\neq0$, then $\Omega_{1}=0$, and so we have
%the polynominal equation%
\begin{equation}
~3H_{P}^{2}-\lambda_{0}-H_{P}^{-n}\lambda_{2}=0. \label{fe.35}%
\end{equation}
Moreover, combining the equation with Eq.(\ref{fe.34}) we obtain $\Omega
_{2}=0$, which implies that the vacuum component $\Lambda\left(  H\right)  $
ensures the existence of de Sitter points. For example, in the simplest
scenario where $n=-1,$ Eq.(\ref{fe.35}) provides the following two critical
points
\[
P_{1}:\left\{  H\left(  P_{1}\right)  ~,~\Omega_{1}\left(  P_{1}\right)
\right\}  =\left\{  \frac{1}{6}\left(  \lambda_{2}-\sqrt{12\lambda_{0}%
+\lambda_{2}^{2}}\right)  ,0\right\}
\]%
\[
P_{2}:\left\{  H\left(  P_{2}\right)  ~,~\Omega_{1}\left(  P_{2}\right)
\right\}  =\left\{  \frac{1}{6}\left(  \lambda_{2}+\sqrt{12\lambda_{0}%
+\lambda_{2}^{2}}\right)  ,0\right\}  .
\]
In order to study the stability of the solution around the critical points
$P_{1},$ and $P_{2}$, we linearize the dynamical system by substituting
\begin{equation}
H\left(  t\right)  =H\left(  P_{\left(  1,2\right)  }\right)  +\varepsilon
H_{\varepsilon}\left(  t\right)  ,~\Omega_{1}\left(  t\right)  =\Omega
_{1}\left(  P_{\left(  1,2\right)  }\right)  +\varepsilon\Omega
_{_{1\varepsilon}}\left(  t\right)
\end{equation}
in the system (\ref{fe.32})-(\ref{fe.33}), with $\varepsilon^{2}\rightarrow0$.
Next, we determine the eigenvalues of the linearized system and conclude that
the point is stable if all the eigenvalues have negative real parts.

\begin{figure}[ptb]
\includegraphics[height=12.5cm]{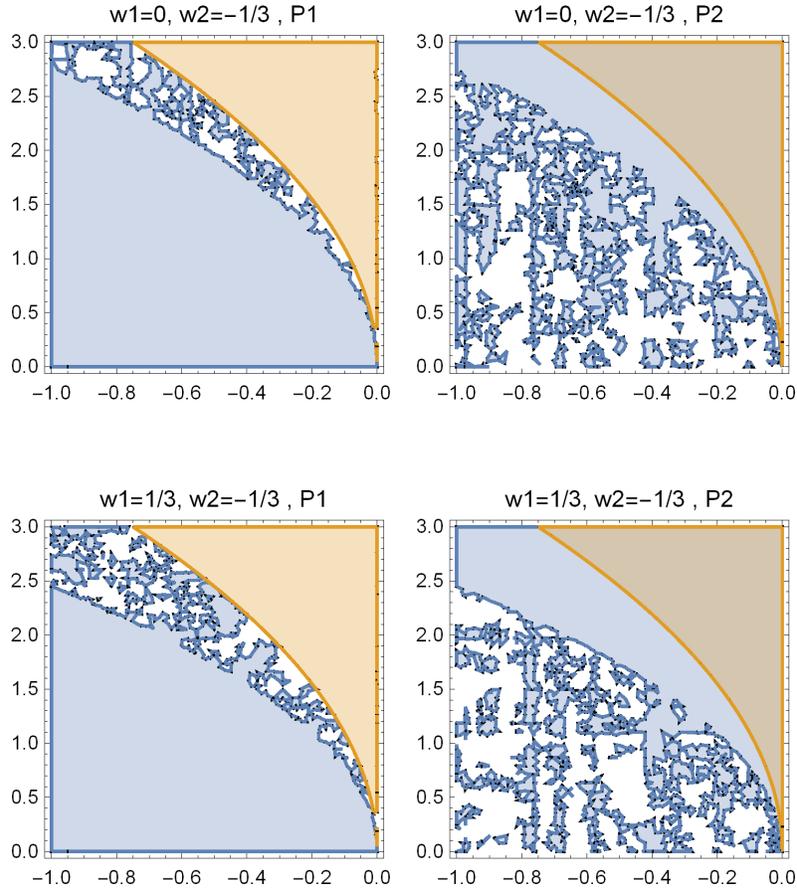}\centering
\caption{Plot of the region in the space $\left\{  \lambda_{0},\lambda
_{2}\right\}  ~$where the eigenvalues of the linearized system (\ref{fe.32}%
)-(\ref{fe.33}) are negative. The first line is for $w_{1}=0$ and $w_{2}%
=\frac{1}{3}$, while the second line is for $w_{1}=\frac{1}{3}$ and $w_{2}=0$.
The plots in the left column corresponds to the point $P_{1}$, and the plots
in the right column to the point $P_{2}$. \ The points are stable only when
the colored areas overlap.}%
\label{image1}%
\end{figure}

\begin{figure}[ptb]
\includegraphics[height=12.5cm]{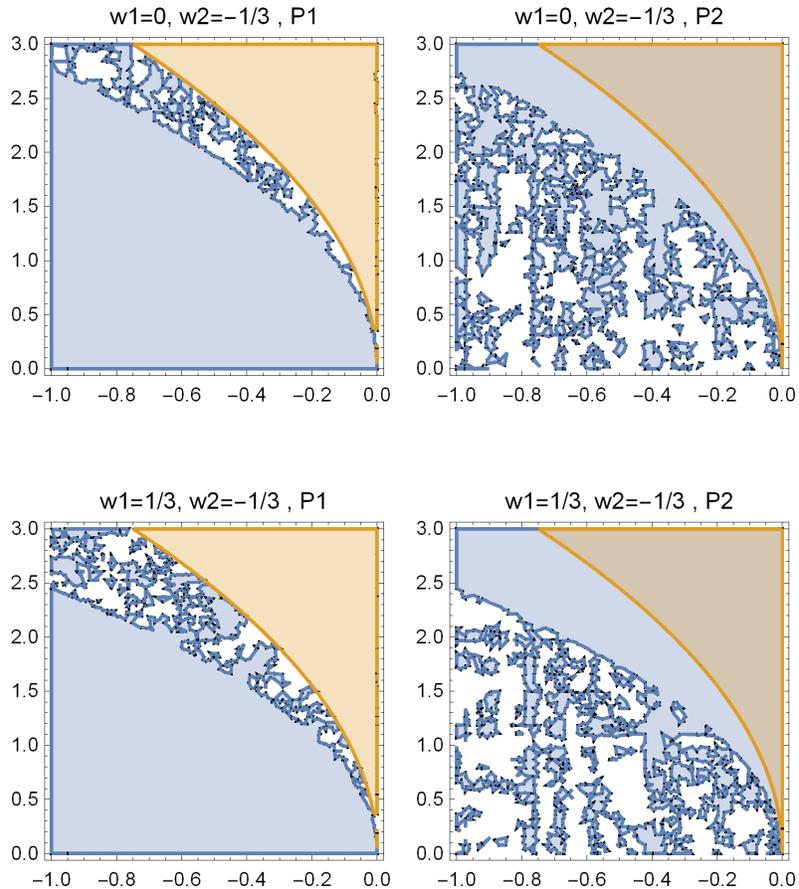}\centering
\caption{Plots of the region in the space $\left\{  \lambda_{0},\lambda
_{2}\right\}  ~$where the eigenvalues of the linearized system (\ref{fe.32}%
)-(\ref{fe.33}) are negative. The first line is for $w_{1}=0$ and
$w_{2}=-\frac{1}{3}$, while the second line is for $w_{1}=\frac{1}{3}$ and
$w_{2}=-\frac{1}{3}$. Recall that $w_{2}=-\frac{1}{3}$ corresponds to the
scenario of a FLRW universe with non-zero spatial curvature. The plots in the
left column corresponds to the point $P_{1}$, and the plots in the right
column to the point $P_{2}$. The points are stable only when the colored areas
overlap.}%
\label{image2a}%
\end{figure}

For $n=-1$, the linearized system around the critical points is%
\begin{equation}
\dot{H}_{\varepsilon}=(\frac{\lambda_{2}}{2}-3)\left(  1+w_{2}\right)
H_{\varepsilon},
\end{equation}%
\begin{equation}
\dot{\Omega}_{1\varepsilon}=\frac{\left(  1+w_{2}\right)  \lambda_{2}}%
{6}H_{\varepsilon}+(3(w_{1}-w_{2})-\left(  1+w_{2}\right)  \lambda_{0}%
)H_{P1}\Omega_{1\varepsilon}.
\end{equation}
The eigenvalues of the linearized system are functions of the parameters
$w_{1},w_{2}$ and $\lambda_{0,}~\lambda_{2}$. In Figs. (\ref{image1}) and
(\ref{image2a}) we present the regions of the parameter space, $\left\{
\lambda_{0},\lambda_{2}\right\}  $, for which the points $P_{1},~P_{2}$ are
real and the eigenvalues have negative real values. The plots are for specific
values of the parameters $w_{1},$ and $w_{2}$. The common area of the regions
provides the appropriate pairs, $\left\{  \lambda_{0},\lambda_{2}\right\}  ,$
for which the de-Sitter points are stable.

Now for $n=1$, Eq. (\ref{fe.35}) \ admits the three critical points
\[
P_{1}:\left\{  H\left(  P_{1}\right)  ~,~\Omega_{1}\left(  P_{1}\right)
\right\}  =\left\{  \frac{1}{3}[(\frac{2\lambda_{0}^{3}}{9\lambda_{2}%
+\sqrt{81\lambda_{2}^{2}-4\lambda_{0}^{3}}})^{1/3}+(\frac{9\lambda_{2}%
+\sqrt{81\lambda_{2}^{2}-4\lambda_{0}^{3}}}{2})^{1/3}],0\right\}
\]%
\[
P_{2}:\left\{  H\left(  P_{2}\right)  ~,~\Omega_{1}\left(  P_{2}\right)
\right\}  =\left\{  \frac{1}{3}[-\frac{(1+\sqrt{-3})\lambda_{0}}{3\sqrt[3]%
{4}(9\lambda_{2}+\sqrt{81\lambda_{2}^{2}-4\lambda_{0}^{3}})^{1/3}}%
+(1-\sqrt{-3})(\frac{9\lambda_{2}+\sqrt{81\lambda_{2}^{2}-4\lambda_{0}^{3}}%
}{432})^{1/3}],0\right\}
\]%
\[
P_{3}:\left\{  H\left(  P_{3}\right)  ~,~\Omega_{1}\left(  P_{3}\right)
\right\}  =\left\{  \frac{1}{3}[-\frac{(1-\sqrt{-3})\lambda_{0}}{3\sqrt[3]%
{4}(9\lambda_{2}+\sqrt{81\lambda_{2}^{2}-4\lambda_{0}^{3}})^{1/3}}%
+(1+\sqrt{-3})(\frac{9\lambda_{2}+\sqrt{81\lambda_{2}^{2}-4\lambda_{0}^{3}}%
}{432})^{1/3}],0\right\}
\]
provided that~$\Delta=12\lambda_{0}^{3}-243\lambda_{2}^{2}\geq0$. Otherwise,
when $\Delta<0$, only $P_{1}$ is a real point and describes always a de Sitter Universe.

Furthermore, for $n=2,~$the algebraic Eq.(\ref{fe.35}) admits the four
solutions,%
\[
P_{1\pm}:\left\{  H\left(  P_{1\pm}\right)  ~,~\Omega_{1}\left(  P_{1\pm
}\right)  \right\}  =\left\{  \pm\sqrt{\frac{1}{6}\left(  \lambda_{0}%
-\sqrt{12\lambda_{2}+\lambda_{0}^{2}}\right)  },0\right\}
\]%
\[
P_{2\pm}:\left\{  H\left(  P_{2\pm}\right)  ~,~\Omega_{1}\left(  P_{2\pm
}\right)  \right\}  =\left\{  \pm\sqrt{\frac{1}{6}\left(  \lambda_{0}%
+\sqrt{12\lambda_{2}+\lambda_{0}^{2}}\right)  },0\right\}  .
\]
which are de Sitter points for values of $\lambda_{0},~\lambda_{2}$ for which
the solutions are real. Easily, it follows that when $\lambda_{0},~\lambda
_{2}$ are positive, then points $P_{2\pm}$ are always real.

When $n=-2$, Eq.(\ref{fe.35}) provides the critical points $P_{1\pm}:$%
\[
\left\{  H\left(  P_{1\pm}\right)  ~,~\Omega_{1}\left(  P_{1\pm}\right)
\right\}  =\left\{  \pm\sqrt{\frac{\lambda_{0}}{3-\lambda_{2}}},0\right\}
~,~\text{with }\left\{  \lambda_{2}<3,\lambda_{0}>0\right\}  \text{ or
}\left\{  \lambda_{2}>0,\lambda_{0}\right\}  .
\]

Finally for $n=-3,$ the cubic equation Eq.(\ref{fe.35}) admits three solutions%

\[
P_{1}:\left\{  H\left(  P_{1}\right)  ~,~\Omega_{1}\left(  P_{1}\right)
\right\}  =\left\{  \frac{1}{\lambda_{2}}-\frac{\sqrt[3]{2}}{\lambda_{2}%
\Sigma\left(  \lambda_{0},\lambda_{2}\right)  }-\frac{\Sigma\left(
\lambda_{0},\lambda_{2}\right)  }{\sqrt[3]{2}\lambda_{2}},0\right\}  ,
\]%
\[
P_{2}:\left\{  H\left(  P_{2}\right)  ~,~\Omega_{1}\left(  P_{2}\right)
\right\}  =\left\{  \frac{1}{\lambda_{2}}+(1+\sqrt{-3})\left(  \frac
{1}{\sqrt[3]{4}\lambda_{2}\Sigma\left(  \lambda_{0},\lambda_{2}\right)
}+\frac{\Sigma\left(  \lambda_{0},\lambda_{2}\right)  }{\sqrt[3]{16}%
\lambda_{2}}\right)  ,0\right\}  ,
\]%
\[
P_{3}:\left\{  H\left(  P_{3}\right)  ~,~\Omega_{1}\left(  P_{2}\right)
\right\}  =\left\{  \frac{1}{\lambda_{2}}-\left(  1-\sqrt{-3}\right)  \left(
\frac{{}}{\sqrt[3]{4}\lambda_{2}\Sigma\left(  \lambda_{0},\lambda_{2}\right)
}-\frac{\Sigma\left(  \lambda_{0},\lambda_{2}\right)  }{\sqrt[3]{16}%
\lambda_{2}}\right)  ,0\right\}  .
\]
where $\Sigma\left(  \lambda_{0},\lambda_{2}\right)  =(-2+\lambda_{0}%
\lambda_{2}^{2}+\sqrt{-4\lambda_{0}\lambda_{2}^{2}+\lambda_{0}^{2}\lambda
_{2}^{4}})^{1/3}$. Point $P_{1}$ is always real, while $P_{2}$ and $P_{3}$
describe de Sitter Universes when $27\lambda_{0}(4-\lambda_{0}\lambda_{2}%
^{2})\geq0.$

\section{Singularities and analytical solutions}

\label{sec4} Technically, in order to study the existence of matter-dominated
eras, that is, singular solutions of the form $a\left(  \tau\right)
=a_{0}\tau^{p},~\tau=t-t_{0}$, we apply the ARS algorithm, hence the
analytical solution can be written as a Painlev\'{e} Series around the
(movable) singularity. Moreover, the ARS algorithm provides important
information regarding the stability of the singular solution $a\left(
\tau\right)  =a_{0}\tau^{p}$. The approach of the ARS algorithm is inspired by
the work of Kowalevskaya \cite{kowa88}, where the solution of a differential
equation is described by a power-law function when the differential equation
admits a movable singularity. Hence, a necessary condition for the approach to
succeed is the existence of at least one movable singularity.\textbf{ }The
position of the singularity is $t_{0}$, and the singulariy is characterized
movable when $t_{0}$ depends on the initial conditions.

The ARS algorithm is developed via a three-step process, a brief description
of which is

Step A: The determination of the leading-order behavior. The leading-order
term needs to be a negative integer, or at least a non-integral rational
number. It is important to point out that the coefficient of the leading-order
term may or may not be determined explicitly.

Step B: The proof of the existence of sufficient number of (arbitrary)
integration constants by determining the resonances (Fuchs indices). The value
minus one should always appear always of one of the resonances. This is
important for the singularity to be movable and hence the ARS algorithm to be valid.

Step C: The consistency test is performed by substituting an expansion of the
power-law series which describes the solution of the original equation, to
test that it is indeed a true solution.

From the leading-order term and the resonances, we can determine the step of
the Laurent expansion (Painlev\'{e} Series) which describes the actual
solution of the original equation. Whereas, the coefficients of the Laurent
expansion are determined from the consistency test. For a Right Painlev\'{e}
Series the resonances must be positive, for a left Painlev\'{e} series the
resonances must be negative, while for a full Laurent expansion the resonances
have to be mixed. Clearly, in the case of an autonomous second-order ordinary
equation, since there is only one free resonance, the possible Laurent
expansions are either left or right Painlev\'{e} series. For more details on
the method we refer the reader to the review article of Ramani et al.
\cite{Ramani 89}.

\subsection{Case A: One ideal fluid}

In this case we consider one ideal fluid. In order to determine the movable
singularities of Eq.(\ref{fr.03c}) we perform the change of variables,
$a\left(  t\right)  \rightarrow A^{-1}\left(  t\right)  $, and multiply the
equation with the term $A^{2}\left(  t\right)  \dot{A}^{n}\left(  t\right)  $
and we find
\begin{equation}
\dot{A}^{n}\left(  2A\ddot{A}-\left(  5+3w_{1}\right)  \dot{A}^{2}+\left(
1+w_{1}\right)  \lambda_{0}A^{2}\right)  -\left(  -1\right)  ^{1-n}\left(
1+w_{1}\right)  \lambda_{2}A^{2+n}=0. \label{fe.09}%
\end{equation}
Then we insert $A\left(  \tau\right)  =A_{0}\tau^{p}$ in the above equation,
where $\tau=t-t_{0}$, and we search for the dominant terms in order to
determine the power $p$. Notice, that $t_{0}$ is an integration constant and
it denotes the position of the singularity.

\subsubsection{Positive power, $n>0$}

For $n>0$ and in the context of $A\left(  \tau\right)  =A_{0}\tau^{p}$
Eq.(\ref{fe.09}) then takes the form%
\begin{equation}
p^{n+1}\left(  3\left(  1+w_{1}\right)  p+2\right)  -p^{n}\tau^{2}\left(
1+w_{1}\right)  \lambda_{0}-\left(  -1\right)  ^{-n}\left(  1+w_{1}\right)
\lambda_{2}\tau^{2+n}=0, \label{fe.10}%
\end{equation}
from which it follows that the only possible dominant term is that of
$p^{n+1}\left[  2+3\left(  1+w_{1}\right)  p\right]  $. Therefore if we assume
that the leading-order behavior describes explicitly the solution at the
singularity then the latter algebraic equation reduces to
\begin{equation}
2+3\left(  1+w_{1}\right)  p\simeq0, \label{fe.11}%
\end{equation}
or
\begin{equation}
p\simeq-\frac{2}{3\left(  1+w_{1}\right)  }. \label{fe.12}%
\end{equation}

Hence, the singular solution of the ideal fluid, without the $\Lambda$-varying
term, describes the actual solution of the model close to the singularity.

For the second step of the ARS algorithm, we substitute $A\left(  \tau\right)
=A_{0}\tau^{-\frac{2}{3\left(  1+w_{1}\right)  }}+m\tau^{-\frac{2}{3\left(
1+w_{1}\right)  }+s}$, and we linearize around $m=0$. From the remaining
terms, we find that the coefficient of the leading-order term vanishes if and
only if
\begin{equation}
s\left(  s+1\right)  =0 \label{fe.13}%
\end{equation}
which means that the resonances are $s_{1}=-1$ and $s_{2}=0$. The second
resonance tells us that the second integration constant is the coefficient
$A_{0}$, of leading-order term. We show that equation (\ref{fe.11}) is not
affected by $A_{0}$, hence this constant is arbitrary. In this case the
analytical solution of Eq.(\ref{fe.09}) is given by a right Laurent expansion,
while its step depends on $w_{1}$. Notice, that the presence of a right
Painlev\'{e} series means that the matter-dominated era close to the
singularity, is an unstable solution\footnote{For a discussion on the relation
between stability of solutions and Laurent expansion see \cite{leachss}.}. We
note that the problem contains two integration constants, namely $A_{0}$ and
$w_{1}$, hence it is not necessary to perform the consistency test. Below, we
complete the current analysis by presenting some analytical solutions which
are physically interesting.

\paragraph{Analytic solution for a non-relativistic fluid:}

For a pressureless matter source, we have $w_{1}=0$. Therefore, from
Eq.(\ref{fe.12}) we calculate $p=-\frac{2}{3}$, which means that the Laurent
expansion has a step of $\frac{1}{3}$, that is, the solution of (\ref{fe.09})
is%
\begin{equation}
A\left(  \tau\right)  =A_{0}\tau^{-\frac{2}{3}}+A_{1}\tau^{-\frac{1}{3}}%
+A_{2}+A_{3}\tau^{\frac{1}{3}}+%
%TCIMACRO{\dsum \limits_{I=4}^{\infty}}%
%BeginExpansion
{\displaystyle\sum\limits_{I=4}^{\infty}}
%EndExpansion
A_{k}\tau^{-\frac{2}{3}+\frac{\nu}{3}}, \label{fe.14}%
\end{equation}
where we need to determine the coefficients $A_{1},A_{2},..$ etc.

In the case of the $\Lambda-$varying model (\ref{fr.03a}), with $n=1$, the
first six non-zero coefficients of (\ref{fe.14}) are: $A_{0}$, which is an
arbitrary integration constant, and%
\[
\frac{A_{6}}{A_{0}}=-\frac{\lambda_{0}}{12},~\frac{A_{9}}{A_{0}}%
=-\frac{\lambda_{2}}{16},~\frac{A_{12}}{A_{0}}=\frac{\left(  \lambda
_{0}\right)  ^{2}}{180},
\]%
\[
\frac{A_{15}}{A_{0}}=\frac{7\lambda_{0}\lambda_{2}}{480}~\text{and~}%
\frac{A_{18}}{A_{0}}=\frac{23895\lambda_{2}^{2}-1072\lambda_{0}^{3}}%
{2903040}A_{0},~etc.,
\]
while the rest of non-zero coefficients are those of $A_{21+3\nu}$,~$\nu\in%
%TCIMACRO{\U{2115} }%
%BeginExpansion
\mathbb{N}
%EndExpansion
.$On the other hand, for other values of the power $n$, the coefficients are
different. For example, in the case of the $\Lambda(H)$ model with $n=3$,
aside from $A_{0},$ the first non-zero coefficients are
\[
\frac{A_{6}}{A_{0}}=-\frac{\lambda_{0}}{12},~\frac{A_{12}}{A_{0}}%
=\frac{\lambda_{0}^{2}}{180},~\frac{A_{15}}{A_{0}}=-\frac{9\lambda_{2}}{160}~
\]%
\[
\frac{A_{18}}{A_{0}}=-\frac{67\lambda_{0}^{2}}{181440},~etc.,
\]
while the other non-zero coefficients are $A_{18+3\nu}$,~$\nu\in%
%TCIMACRO{\U{2115} }%
%BeginExpansion
\mathbb{N}
%EndExpansion
.$

From the aforementioned solutions, we observe that the solution close to the
matter-dominated era includes an additional term $A_{0}\tau^{\frac{4}{3}}$,
namely
\begin{equation}
A\left(  \tau\right)  \simeq A_{0}\tau^{-\frac{2}{3}}+A_{6}\tau^{\frac{4}{3}}
\label{fe.15}%
\end{equation}
and so the scale factor is written as a Taylor expansion around $\tau=0$ as
follows,
\begin{equation}
a\left(  \tau\right)  \simeq a_{0}\tau^{\frac{2}{3}}+a_{6}\tau^{\frac{8}{3}}.
\label{fe.16}%
\end{equation}

\paragraph{Analytic solution for a radiation fluid:}

Considering that the ideal fluid is radiation ($w_{1}=1/3$) from
(\ref{fe.12}), we obtain $p=-\frac{1}{2}$, which means that the solution is
expressed by a right Painlev\'{e} series with step $\frac{1}{2}$, that is,%
\begin{equation}
A\left(  \tau\right)  =\bar{A}_{0}\tau^{-\frac{1}{2}}+\bar{A}_{1}+\bar{A}%
_{2}\tau^{\frac{1}{2}}+\bar{A}_{3}\tau+%
%TCIMACRO{\dsum \limits_{I=4}^{\infty}}%
%BeginExpansion
{\displaystyle\sum\limits_{I=4}^{\infty}}
%EndExpansion
\bar{A}_{I}\tau^{-\frac{1}{2}+\frac{\nu}{2}}, \label{fe.17}%
\end{equation}
where $\bar{A}_{0}$ is an integration constant.

For $n=1$, the nonzero coefficients of the expansion (\ref{fe.17}) are%
\[
\frac{\bar{A}_{4}}{\bar{A}_{0}}=-\frac{\lambda_{0}}{9}~,~\frac{\bar{A}_{6}%
}{\bar{A}_{0}}=-\frac{\lambda_{2}}{9}~\ ,~\frac{\bar{A}_{8}}{\bar{A}_{0}%
}=\frac{\lambda_{0}^{2}}{90}~,~\frac{~\bar{A}_{10}}{\bar{A}_{0}}%
=\frac{17\lambda_{0}\lambda_{2}}{405},
\]%
\[
\frac{\bar{A}_{12}}{\bar{A}_{0}}=\frac{1665\lambda_{2}^{2}-61\lambda_{0}^{3}%
}{51030},~~\frac{\bar{A}_{14}}{\bar{A}_{0}}=-\frac{221\lambda_{0}^{2}%
\lambda_{2}}{17010}.
%~,~\frac{\bar{A}_{17+2\nu}}{\bar{A}_{0}}=f\left(  \Lambda_{0},\Lambda
%_{1}\right)  ,~\nu\in%
%TCIMACRO{\U{2115} }%
%BeginExpansion
\mathbb{N}
%EndExpansion
.
\]

For $n=2$, the coefficients of the expansion are%
\[
\frac{\bar{A}_{4}}{\bar{A}_{0}}=-\frac{\lambda_{0}}{9}~,~\frac{\bar{A}_{8}%
}{\bar{A}_{0}}=\frac{\lambda_{0}^{2}-12\lambda_{2}}{90}~,~\frac{\bar{A}_{12}%
}{\bar{A}_{0}}=\frac{4212\lambda_{0}\lambda_{2}-61\left(  \lambda_{0}\right)
^{3}}{51030},
\]%
\[
\frac{\bar{A}_{16}}{\bar{A}_{0}}=\frac{1261\lambda_{0}^{4}-362664\lambda
_{0}^{4}\lambda_{2}+879984\lambda_{2}^{2}}{9185400}
%~,~\frac{\bar{A}_{20+4\nu}}{\bar{A}_{0}%
%}=g\left(  \Lambda_{0},\Lambda_{1}\right)  ~,~\nu\in%
%TCIMACRO{\U{2115} }%
%BeginExpansion
\mathbb{N}
%EndExpansion
.
\]

Therefore, the analytical solution prior to the radiation era is well
approximated by the following expression:
\[
A\left(  \tau\right)  =\bar{A}_{0}\tau^{-\frac{1}{2}}\left(  1-\frac
{\lambda_{0}}{9}\tau^{2}\right)  ,
\]
and so
\begin{equation}
a\left(  \tau\right)  \simeq a_{0}\tau^{\frac{1}{2}}+a_{1}\tau^{\frac{5}{2}}.
\label{fe.18}%
\end{equation}

\subsubsection{Negative power, $n<0$}

Now we study the case where the exponent $n$ in $\Lambda(H)=\lambda
_{0}+\lambda_{2}H^{-n}$ is negative. We remind the reader that we are still
using the variable $A\left(  \tau\right)  =A_{0}\tau^{p}$. Under those
conditions Eq.(\ref{fe.09}) gives
\begin{equation}
\tau^{\left(  2-n\right)  \left(  p-1\right)  }\left[  \left(  3\left(
1+w_{1}\right)  p+2\right)  p\tau^{-n}-\left(  1+w_{1}\right)  \lambda_{0}%
\tau^{2-n}-\left(  -1\right)  ^{-n}\tau p^{-n}\left(  1+w_{1}\right)
\lambda_{2}\tau^{2}\right]  =0. \label{fe.19}%
\end{equation}
Following similar notations to those of section 4.1.1 we find that the only
possible leading term is the one with coefficient $2+3\left(  1+w_{1}\right)
p$ as long as $n>-2$, which implies that the exponent $p$ is given by
Eq.(\ref{fe.12}). As expected, from the ARS algorithm, we find that the
constant $A_{0}$ is arbitrary, hence the corresponding resonances
(\ref{fe.13}) are $s_{1}=-1$ and $s_{2}=0$ respectively, whilst the analytical
solution is written as a right Painlev\'{e} series. Specifically, we find the
following interesting solutions.

\paragraph{Analytic solution for a non-relativistic fluid:}

For $w_{1}=0,$ the Laurent expansion is of the form of (\ref{fe.14}), where
with $n=-1$, namely $\Lambda(H)=\lambda_{0}+\lambda_{2}H$, we obtain the
following coefficients:
\[
\frac{A_{3}}{A_{0}}=-\frac{\lambda_{2}}{6}~,~\frac{A_{6}}{A_{0}}=\frac
{\lambda_{2}^{2}-12\lambda_{0}}{144}~,~\frac{A_{9}}{A_{0}}=\frac{\lambda
_{2}^{3}+36\lambda_{0}\lambda_{2}}{2592}~,
\]%
\[
\frac{A_{12}}{A_{0}}=\frac{216\lambda_{0}^{2}-9\lambda_{0}\lambda_{2}%
^{2}-\lambda_{2}^{3}}{38880}~.
%\frac{A_{15+3\nu}}{A_{0}}=h\left(  \Lambda_{0},\Lambda_{1}\right)
%,~\nu\in%
%TCIMACRO{\U{2115} }%
%BeginExpansion
\mathbb{N}
%EndExpansion
\]
This means that close to the singularity the scale factor is approximated by
\begin{equation}
a\left(  \tau\right)  \simeq a_{0}\tau^{\frac{2}{3}}+a_{3}\tau^{\frac{5}{3}}.
\label{fe.20}%
\end{equation}
Notice that the second term of this solution is different with that of
Eq.(\ref{fe.16}) for $n>0$. The closed-form solution in the case of the
$\Lambda(H)=\lambda_{0}+\lambda_{2}H$ running-vacuum model has been found
earlier in \cite{gomez}. In this case the running vacuum model introduces a
mild dynamical behavior of the vacuum energy at low energies, namely dark
matter and dark energy eras. For more details regarding the dynamics as well
as the corresponding observational constraints of the current vacuum model we
refer the reader the work of \cite{gomez}. On the other hand, terms of the
form $H^{3}$, $H^{4}$ ($n<-2$) they can be important for the early universe
(see \cite{perico} and \cite{Sahni2014}).

\paragraph{Analytic solution for a radiation fluid:}

Here, the equation of state parameter (hereafter EoS) is $w_{1}=\frac{1}{3}$,
and the analytical solution is given by the series (\ref{fe.17}). For $n=-1,$
the corresponding non-zero coefficients are
\[
\frac{\bar{A}_{2}}{A_{0}}=-\frac{\lambda_{2}}{6},~\frac{\bar{A}_{4+2\nu}%
}{A_{0}}\neq0
%\bar{h}\left(  \Lambda_{0},\Lambda_{1}\right)
~\nu\in%
%TCIMACRO{\U{2115} }%
%BeginExpansion
\mathbb{N}
%EndExpansion
,
\]
and so, prior to the radiation-dominated era (singular solution), the scale
factor can be written as
\begin{equation}
a\left(  \tau\right)  \simeq a_{0}\tau^{\frac{1}{2}}+a_{2}\tau^{\frac{3}{2}}.
\label{fe.21}%
\end{equation}
Again, this solution includes an extra term which differs from Eq.(\ref{fe.18}%
) for $n>0$.

\begin{figure}[ptb]
\includegraphics[height=6cm]{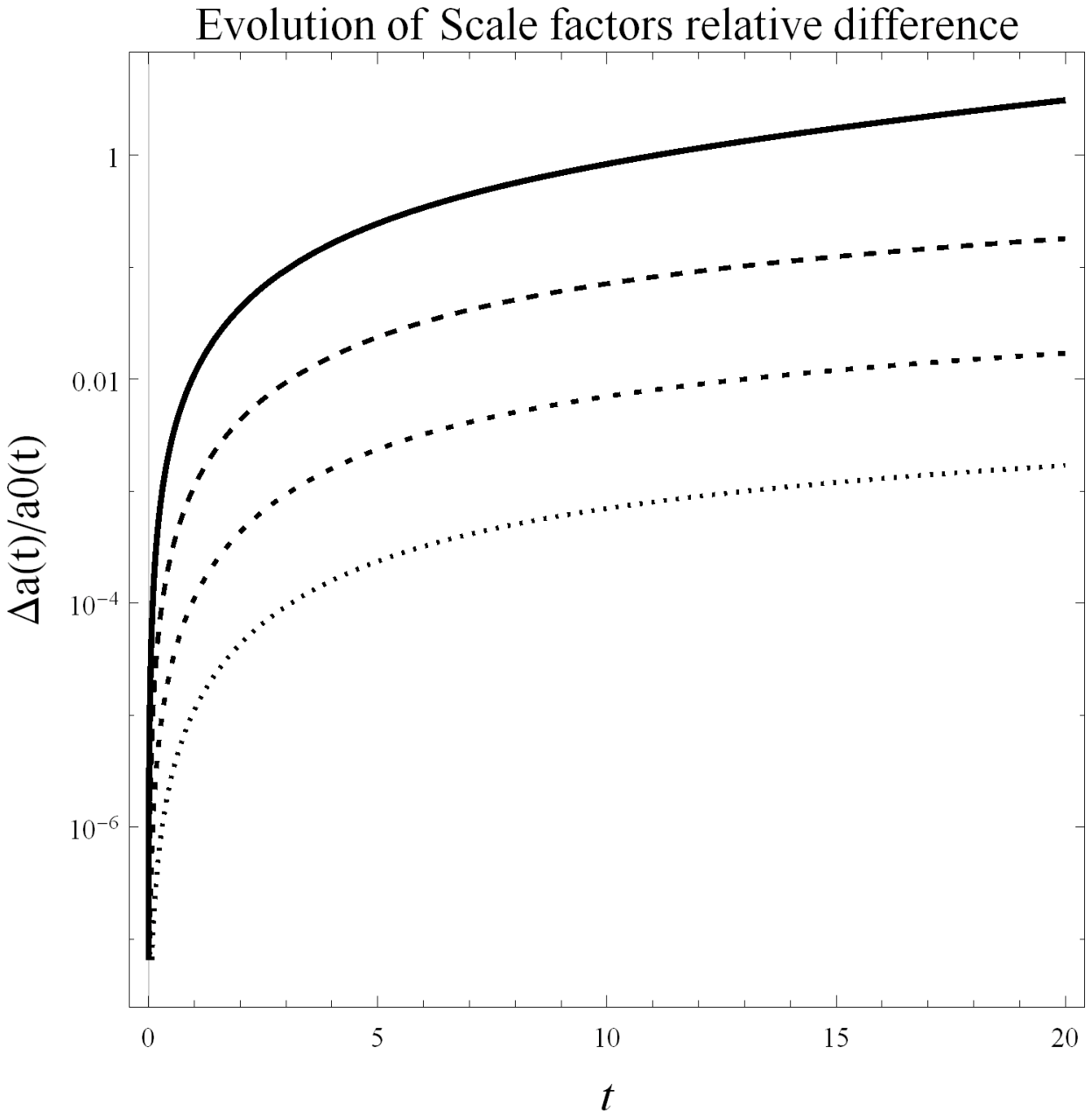}\centering
\includegraphics[height=6cm]{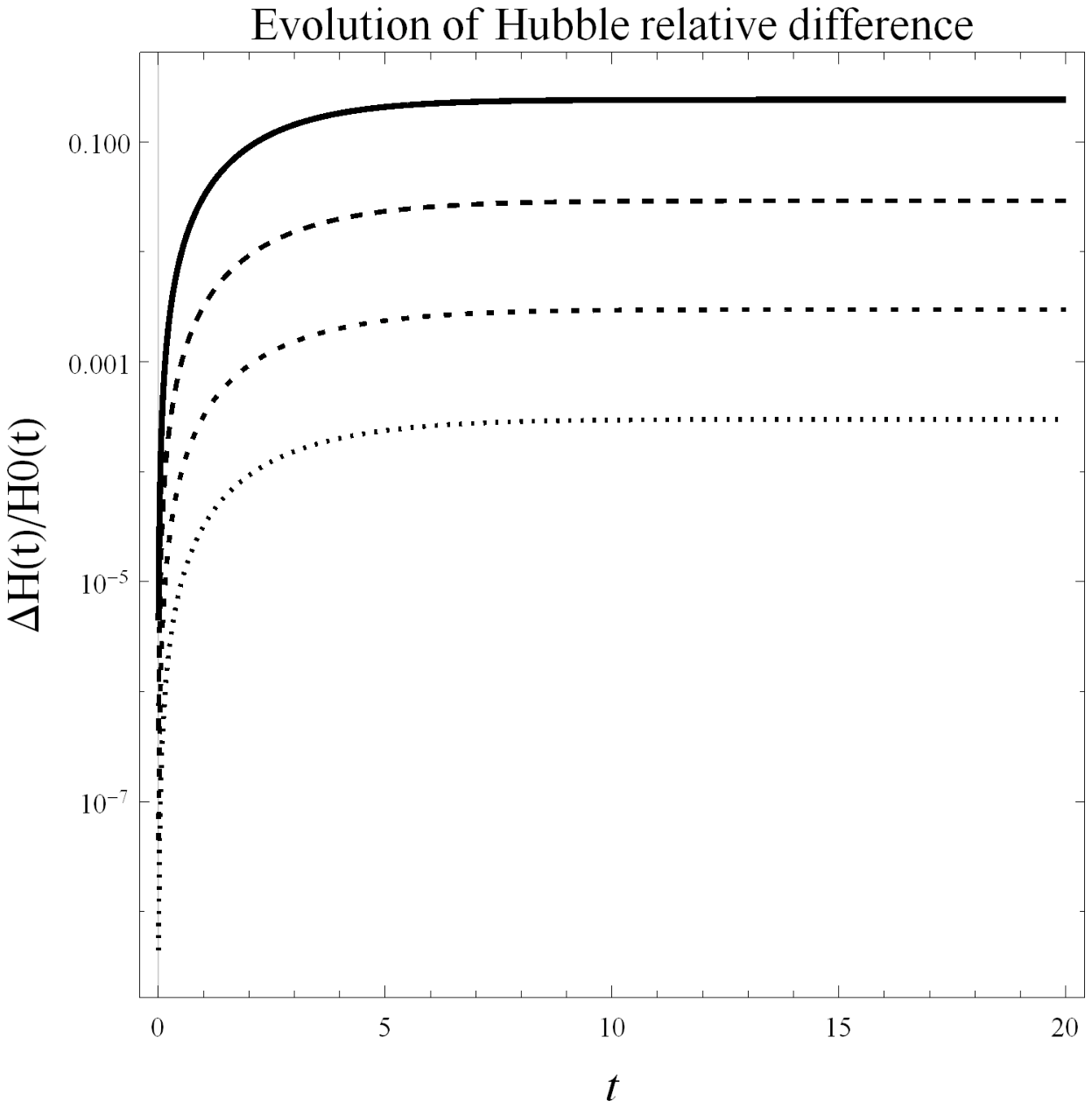}\centering
\caption{Qualitative evolution in time for the relative difference between the
solutions of the master equation (\ref{fr.03b}) for $\lambda_{0}=0$, with
$\lambda_{0}\neq0$ and $n=-1$. Left Fig. is for the scale factors while Right
Fig. for the Hubble functions. For the numerical integration we considered
$\lambda_{2}=1$, while the solid line provides the relative error of solution
with $\lambda_{0}=10^{-1}\lambda_{2},~$dash-dash line for $\lambda_{0}%
=10^{-2}\lambda_{l}$, dot-dot line for $\lambda_{0}=10^{-3}\lambda_{2}$ and
the dash-dot line for $\lambda_{0}=10^{-4}\lambda_{2}$. For the numerical
integration we started close to the singular solution with $w_{1}=\frac{1}{3}%
$, that is, at the radiation epoch. }%
\label{relerror}%
\end{figure}

At this point the following comments are appropriate. The current approximate
solutions (\ref{fe.20})-(\ref{fe.21}), for non-relativistic matter and
radiation contain extra terms which differ from those in (\ref{fe.16}) and
(\ref{fe.18}) where $n=1$. This is expected: for negative powers of $n$, the
extra terms of the solutions are related to the $\lambda_{2}H^{-n}$ term in
Eq.(\ref{fr.03a}), while for $n>0$ the extra terms are affected by the
constant vacuum term $\lambda_{0}$. However, in this case the $\lambda_{0}$
term is introduced in the second-order correction to the solution, which means
that there are differences between the two solutions $\lambda_{0}=0$ and
$\lambda_{0}\neq0$, not very far from the singularity. In order to understand
the differences between the two solutions, in Fig. \ref{relerror} we present
the evolution of the relative difference between the numerical simulations for
$n=-1$, and $\lambda_{0}=0$ and those with $\lambda_{0}\neq0$, for the scale
factor and the Hubble expansion rate.

\begin{figure}[ptb]
\includegraphics[height=6cm]{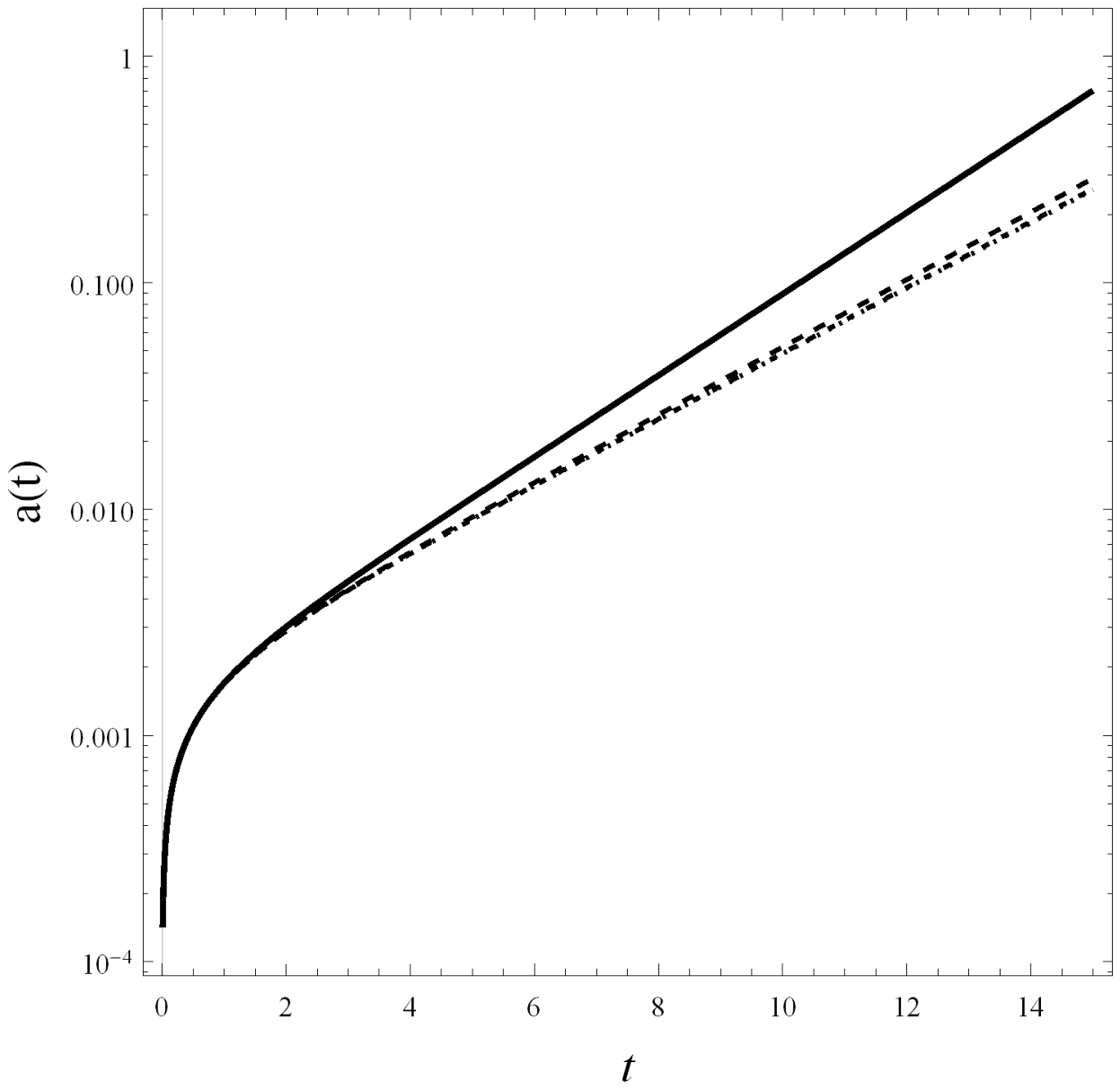}\centering
\includegraphics[height=6cm]{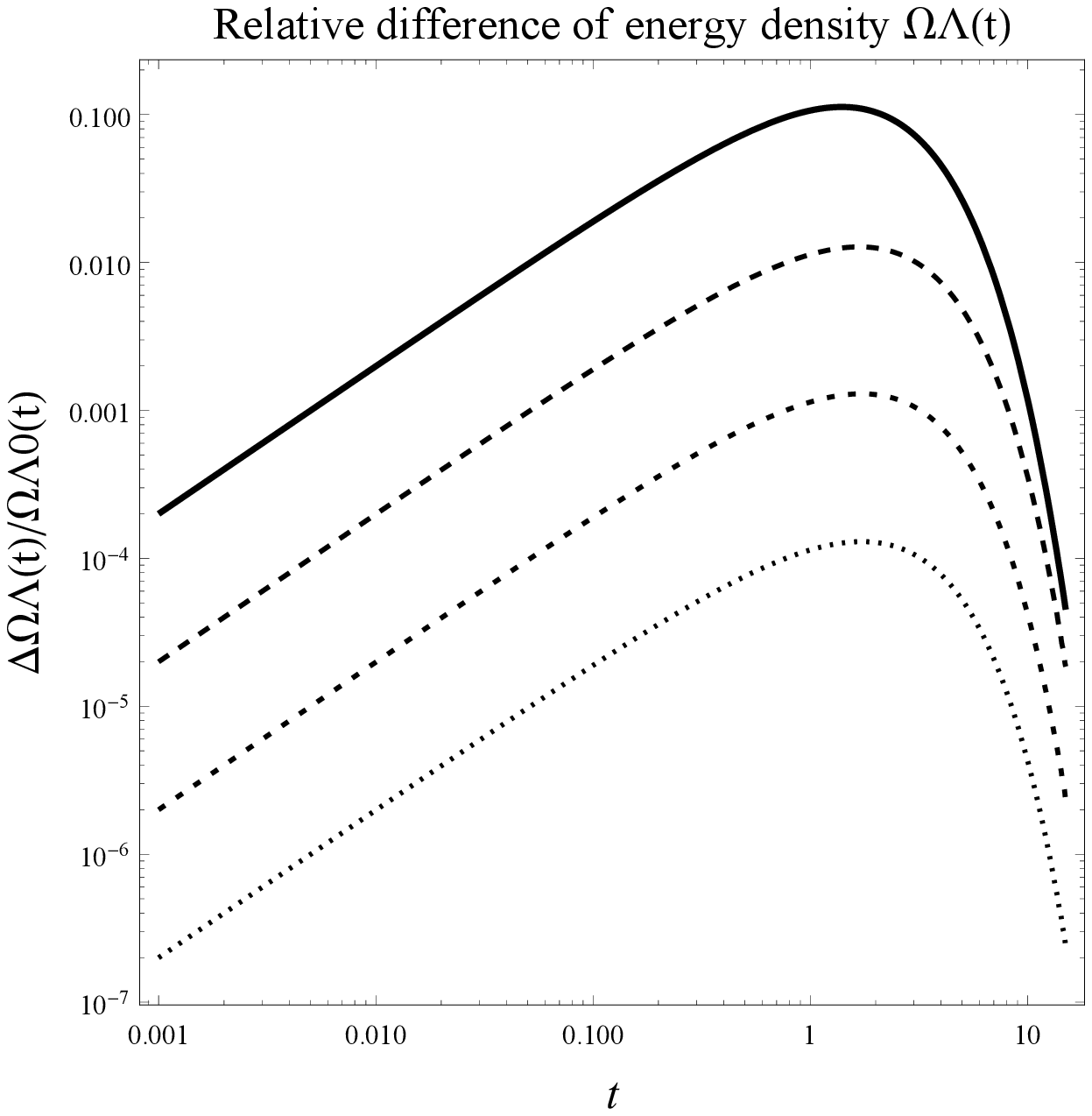} \centering\caption{Qualitative
evolution in time for the scale factor $a(t)$ (Left Fig.) and the relative
difference of the energy density $\Omega_{\Lambda}\left(  t\right)  $ for the
solutions which were presented in Fig. (\ref{relerror}). For the numerical
integration we have considered $\lambda_{2}=1$, $n=-1$, while the solid line
provides the relative error of the solution with $\lambda_{0}=10^{-1}%
\lambda_{2},~$the dash-dash line for $\lambda_{0}=10^{-2}\lambda_{2}$, the
dot-dot line for $\lambda_{0}=10^{-3}\lambda_{2}$ and the dash-dot line for
$\lambda_{0}=10^{-4}\lambda_{2}$. We started the numerical integration close
to the singular solution with $w_{1}=\frac{1}{3}$, that is, at the radiation
epoch. }%
\label{scale1}%
\end{figure}

We observe that prior to the singularity the relative error is close to the
error of the numerical integration; however, as the system evolves the
relative error becomes larger. In the left panel of Fig. \ref{scale1} we plot
the evolution of the scale factors which are considered in Fig. \ref{relerror}%
, while in the right panel of Fig. \ref{scale1} we show the relative
difference in the energy density $\Omega_{\Lambda}=\Lambda\left(  H\right)
/3H^{2}$ between the different solutions.

Recall that $\lambda_{0}$ does not correspond to the value of the cosmological
constant at the present time. In particular, the latter is given by
$\Lambda\left(  H_{0}\right)  =\lambda_{0}+\lambda_{2}H_{0}^{-n}$, where
$H_{0}$ is the value of the Hubble function at the present time. Therefore,
since we know that $\Lambda\left(  H_{0}\right)  $ is small, we must have
$\lambda_{0}$ of the order of $\lambda_{2}H_{0}^{-n}$: that is, for $n<0$,
$\left\vert \lambda_{0}\right\vert >\left\vert \lambda_{2}\right\vert .$ This
makes clear that the correction terms which depend on $\lambda_{0}$ are
important for the evolution of the system.

The singularity analysis fails for $n<-2$, and below we will carry out our
analysis for the case of two ideal fluids.

\subsection{Case B: Two ideal gases}

In this section we consider that the cosmic fluid consists two ideal fluids,
where one of them is minimally coupled (see Case B in section 2). In this case
using the transformation $A\left(  t\right)  =a^{-1}\left(  t\right)  $, the
second-order differential equation (\ref{fe.08}) becomes
\begin{equation}
\dot{A}^{n}[2A\ddot{A}-\left(  5+3w_{1}\right)  \dot{A}^{2}+\left(
1+w_{1}\right)  \lambda_{0}A^{2}+w_{2}\rho_{20}A^{3\left(  1+w_{2}\right)
}]-\left(  -1\right)  ^{-n}\left(  1+w_{1}\right)  \lambda_{2}A^{2-2n}=0.
\label{fe.22}%
\end{equation}
Using $\tau=t-t_{0}$ and substituting $A\left(  t\right)  =A_{0}\tau^{p}$ in
(\ref{fe.22}), we find the expression
\begin{equation}
\tau^{-3+\left(  3+n\right)  p}\left[  p^{n+1}\left(  3\left(  1+w_{1}\right)
p-2\right)  -p^{n}\tau^{2}\left(  \left(  1+w_{1}\right)  \lambda_{2}%
-w_{2}\rho_{20}A_{0}^{3\left(  1+w_{2}\right)  }\tau^{3\left(  1+w_{2}\right)
p}\right)  -\left(  -1\right)  ^{n}\left(  1+w_{1}\right)  \lambda_{2}%
\tau^{2+n}\right]  =0. \label{fe.23}%
\end{equation}
This provides the following three dominant behaviors:

\subsubsection{Positive power, $n>0,$ solution I}

For $n>0$, if we assume that the dominant term close to the singularity is
$2+3(1+w_{1})p$, then $p$ is given by Eq.(\ref{fe.12}). This implies that the
singular solution describes the matter-dominated era via the fluid $\rho
_{1}\left(  t\right)  $. Of course we need to clarify that the above term is
indeed the dominant one if and only if the corresponding EoS parameters of the
two ideal fluids satisfy the inequality
\begin{equation}
w_{2}<w_{1}. \label{fe.24}%
\end{equation}

Within the framework of this latter restriction we present some special
analytical solutions, below. It is important to mention that when
$w_{2}=-\frac{1}{3}$, Eq.(\ref{fe.08}) corresponds to that of one ideal
effective fluid in a FLRW spacetime with nonzero spatial curvature $k$, such
that $k=-\rho_{20}$. \emph{ }

For the calculation of the resonances, we follow standard steps; namely, we
substitute $A\left(  t\right)  =A_{0}\tau^{-\frac{2}{3\left(  1+w_{1}\right)
}}\left(  1+mt^{s}\right)  $ in (\ref{fe.22}) and from the coefficient of the
dominant term in the linearized equation around the value $m=0$ we obtain
$s\left(  1+s\right)  =0$, hence the two resonances are $-1$ and $0$. \ 

Because the second resonance has the value zero, the second integration
constant is the coefficient of the dominant term; hence, it is not necessary
to perform the consistency test. However, in order to compare the current
solutions with those of Case A we shall now investigate some applications.

\paragraph{Analytic solution for a non-relativistic fluid and curvature term:}

Here, we have $(w_{1},w_{2})=(0,-\frac{1}{3})$. Therefore, the leading-order
term is $p=-\frac{2}{3}$ and the step of the expansion is $\frac{1}{3}$. The
analytical solution is given by expression (\ref{fe.14}) while, for $n=1,$ the
nonzero coefficients of the Laurent expansion
%have the form $A_{10+2\nu}$ and
%they
are given by
\[
\frac{A_{2}}{A_{0}}=-\frac{3A_{0}^{2}\rho_{20}}{20}~,~\frac{A_{4}}{A_{0}%
}=\frac{9\left(  A_{0}^{2}\rho_{20}\right)  ^{2}}{280}~,~\frac{A_{6}}{A_{0}%
}=-\frac{100\lambda_{0}+9\left(  A_{0}^{2}\rho_{20}\right)  ^{3}}{1200},
\]%
\[
\frac{A_{8}}{A_{0}}=\frac{9\left(  39200\lambda_{0}+3483\left(  A_{0}^{2}%
\rho_{20}\right)  ^{3}\right)  \left(  A_{0}^{2}\rho_{20}\right)  }{17248000}
%~,~\frac{A_{_{10+2\nu}}}{A_{0}}=f\left(
%\lambda_{0},\lambda_{2},\left(  A_{0}\right)  ^{2}\rho_{20}\right)  ~,~\nu\in%
%TCIMACRO{\U{2115} }%
%BeginExpansion
\mathbb{N}
%EndExpansion
.
\]
%have the form $A_{10+2\nu}$ and
%they
Notice that the other non-zero coefficients of the series have the form
%while the other non-zero coefficients are
$A_{10+2\nu}$,~$\nu\in%
%TCIMACRO{\U{2115} }%
%BeginExpansion
\mathbb{N}
%EndExpansion
.$ Recall that the constraint (\ref{fe.01}) provides an algebraic relation
between the free parameters of the model. As expected, for $k=\rho_{20}=0$ the
current solution reduces to that of section 3.1, while close to the
singularity, and for $k=\rho_{20}\neq0,$ the approximated solution is
\begin{equation}
A\left(  \tau\right)  =A_{0}\tau^{-\frac{2}{3}}+A_{2}, \label{fe.25}%
\end{equation}
or%
\begin{equation}
a\left(  \tau\right)  =a_{1}\tau^{\frac{2}{3}}+a_{2}\tau^{\frac{4}{3}}.
\label{fe.26}%
\end{equation}

\paragraph{Analytic solution for radiation and non-relativistic matter:}

Now, we assume that $\rho_{1}\left(  t\right)  $ and $\rho_{2}\left(
t\right)  $ are the radiation ($w_{1}=\frac{1}{3}$) and matter ($w_{2}=0$)
densities respectively. In this case, the solution of the problem are the
series (\ref{fe.17}) with $\bar{A}_{0}$ arbitrary and for $n=1$ the nonzero
coefficients are found to be
\[
\frac{\bar{A}_{4}}{\bar{A}_{0}}=-\frac{\lambda_{0}}{9}~,~\frac{\bar{A}_{6}%
}{\bar{A}_{0}}=-\frac{\lambda_{2}}{9}~,~\frac{\bar{A}_{8}}{\bar{A}_{0}}%
=\frac{\lambda_{0}^{2}}{90}
%,~\frac{\bar{A}_{10+2\nu}}{\bar
%{A}_{0}}\ne 0 g\left(  \Lambda_{0},\Lambda_{1}\right)  ~,~\nu\in%
%TCIMACRO{\U{2115} }%
%BeginExpansion
\mathbb{N}
%EndExpansion
,
\]
%have the form $A_{10+2\nu}$ and
%they
%Notice that the other non-zero coefficients of the series
%have the form
while the other non-zero coefficients take the form $A_{10+2\nu}$,~$\nu\in%
%TCIMACRO{\U{2115} }%
%BeginExpansion
\mathbb{N}
%EndExpansion
.$

\paragraph{Analytic solution for radiation and curvature term:}

Here we consider that the cosmic fluid contains radiation ($w_{1}=\frac{1}{3}%
$) and curvature ($w_{2}=-\frac{1}{3}$). Again, if in the case of the dust we
consider another fluid, for instance $w_{2}=-\frac{1}{3}$, which corresponds
to the curvature term, then the analytical solution is given by
Eq.(\ref{fe.17}) with the following nonzero coefficients (for $n=1$),
\[
\frac{\bar{A}_{2}}{\bar{A}_{0}}=-\frac{A_{0}^{2}\rho_{20}}{12}~,~\frac{\bar
{A}_{4}}{\bar{A}_{0}}=-\frac{32\lambda_{0}-3\left(  A_{0}^{2}\rho_{20}\right)
^{2}}{288}~,
\]%
\[
\frac{\bar{A}_{6}}{\bar{A}_{0}}=\frac{64\lambda_{0}A_{0}^{2}\rho_{20}-5\left(
A_{0}^{2}\rho_{20}\right)  ^{3}-384\lambda_{2}}{3456}
%~,~~\frac{\bar{A}_{8+2\nu}}{\bar{A}_{0}}=g\left(
%\Lambda_{0},\Lambda_{1},\left(  A_{0}\right)  ^{2}\rho_{20}\right)  ~,~\nu\in%
%TCIMACRO{\U{2115} }%
%BeginExpansion
\mathbb{N}
%EndExpansion
,
\]
where we clearly observe the differences with respect to that of the previous
solution (radiation and non-relativistic matter).

\subsubsection{Positive power, $n>0,$ solution II}

The second class of solutions are those for which the leading-order terms of
Eq. (\ref{fe.23}) are $-3+\left(  3+n\right)  p$ and~$-1+p\left[  3\left(
2+w_{2}\right)  +n\right]  $. Keeping only these terms, Eq. (\ref{fe.23})
reduces to
\begin{equation}
-3+\left(  3+n\right)  p\simeq-1+p\left[  3\left(  2+w_{2}\right)  +n\right]
, \label{fe.27}%
\end{equation}
and so
\begin{equation}
p\simeq-\frac{2}{3\left(  1+w_{2}\right)  }. \label{fe.28}%
\end{equation}
Therefore, that new singular solution describes the matter-dominated era of
the minimally coupled fluid $\rho_{2}\left(  t\right)  $.

In contrast to the aforementioned cases, the constant $A_{0}$ here is not
arbitrary but it is given in terms of $w_{1}$, $w_{2}$ and $\rho_{20}$, by
\begin{equation}
\rho_{20}=\frac{4A_{0}^{-3\left(  1+w_{2}\right)  }\left(  w_{2}-w_{1}\right)
}{3w_{2}\left(  1+w_{2}\right)  ^{2}}, \label{fe.29}%
\end{equation}
where $w_{2}\neq0$. Since $A_{0}$ is not arbitrary, we expect that the second
resonance cannot be zero. Indeed, using our methodology we find the following resonances:%

\begin{equation}
s_{1}=-1~~,~~s_{2}=2\frac{\left(  w_{2}-w_{1}\right)  }{1+w_{2}}.
\label{fe.30}%
\end{equation}
Unlike the previous cases, where the corresponding solutions are written as
right Painlev\'{e} series, we have a different situation here. In particular,
if we select the EoS parameters $(w_{1},w_{2}$) such that $s_{2}<0,$ then the
solution is given by a left Painlev\'{e} series and thus the dominant term is
an attractor.\footnote{For more details, we refer the reader to
\cite{bar1,bar2,bar3} and references therein.}

%In contrast to the previous where the solution is expressed by a Right
%Painlev\'{e} Series which indicates the instability of the dominant term, here
%for values of the parameters $w_{1},~w_{2}$ such that $s_{2}<0$, the solution
%is given in terms of\ a Left Painlev\'{e} Series, so that the dominant term is
%an attractor, while in that cases there exists a natural barrier around the
%singularity
%\footnote{For more details, we refer the reader to
%\cite{bar1,bar2,bar3} and references therein.}.

\paragraph{Analytic solution for radiation and curvature:}

Using $w_{1}=\frac{1}{3}$ and $w_{2}=-\frac{1}{3}$ Eqs.(\ref{fe.28}) and
(\ref{fe.30}) provide $p=-1$ and $s_{1}=-1$,~$s_{2}=1$. Therefore, the
analytical solution is
\begin{equation}
A\left(  \tau\right)  =A_{0}\tau^{-1}+A_{1}+A_{2}\tau+A_{3}\tau^{2}+%
%TCIMACRO{\dsum \limits_{v=4}^{\infty}}%
%BeginExpansion
{\displaystyle\sum\limits_{v=4}^{\infty}}
%EndExpansion
A_{v}\tau^{-1+\nu}. \label{fe.31}%
\end{equation}
Notice, that the solution passes the consistency test. Moreover, the fact that
the second resonance is positive implies that the solution (\ref{fe.31}) is
unstable. Now, using the dominant term $A\simeq A_{0}\tau^{-1}$ of the above
solution, i.e. $a(\tau)\propto\tau,~$we find that the Milne universe and the
$R_{h}=ct$ model \cite{Melia} can be seen as perturbations around the present
%we find that the current non-flat
$\Lambda(H)$ model.
%can be seen as equivalent cosmologies as far as the
%cosmic expansion is concerned.

%, despite the fact that the
%two models live in a completely different physical background.
If the $R_{h}=ct$ or Milne models were somehow the
%late-time
effective behavior of the current $\Lambda(H)$ model, then we could argue that
the $R_{h}=ct$ and Milne models are both unstable.
%it is worth to mention that Another important
%information that we can recover is that because of the positive value for the
%second resonance, Milne universe in unstable.

To this end, for $n=1$ the nonzero coefficients of (\ref{fe.31}) are, $A_{1}$
which is the second integration constant and%
\[
A_{2}=\frac{27A_{1}^{2}-2A_{0}^{2}\lambda_{0}}{18A_{0}}~,~A_{3}=\frac
{90A_{1}^{3}-8A_{0}^{2}A_{1}\lambda_{0}-3A_{0}^{3}\lambda_{2}}{36A_{0}^{2}%
}~,~etc,
\]
where $A_{1}$ is a second integration constant.

\subsubsection{Negative power, $n<0$}

The last case that we have to consider is when the power $n$ is negative, i.e.
$n<0$. In this case it follows that the only possible dominant behavior is
that with~$p=-\frac{2}{3\left(  1+w_{1}\right)  }$, while the corresponding
resonances are $s_{1}=-1$ and $s_{2}=0$, which means that the second
integration constant is the $A_{0}$ coefficient. Thus, since we know the two
integration constants explicitly it is not necessary to perform step C of the
ARS algorithm, i.e. the consistency test. However, for the completeness of our
analysis we present the Laurent expansions for two cases of special interest.

\paragraph{Analytic solution for radiation and nonrelativistic matter}

If we assume that $\ w_{1}=\frac{1}{3}$ and $w_{2}=0$, then the dominant
behavior is $p=-\frac{1}{2}$, and the solution is given as a right
Painlev\'{e} series with step $\frac{1}{2}$, that is, series (\ref{fe.17}),
where now the first nonzero coefficients for $n=-1$ are%
\[
\frac{\bar{A}_{2}}{A_{0}}=-\frac{\lambda_{2}}{6}~,~\frac{\bar{A}_{4}}{A_{0}%
}=\frac{\lambda_{2}^{2}-24\lambda_{0}}{216}~,~A_{6}=\frac{\lambda_{2}%
^{3}+24\lambda_{0}\lambda_{2}}{1296},~etc.
\]

\paragraph{Analytic solution for radiation and curvature}

In a similar way, for $w_{1}=\frac{1}{3}$ and $w_{2}=-\frac{1}{3}$, the
solution is given by the right Painlev\'{e} series (\ref{fe.17}) with step
$\frac{1}{2}$, where the\ first nonzero coefficients for $n=-1$ are%
\[
\frac{\bar{A}_{2}}{A_{0}}=-\frac{2\lambda_{2}+A_{0}^{2}\rho_{20}}{12}%
~,~\frac{\bar{A}_{4}}{A_{0}}=\frac{4\lambda_{2}^{2}-96\lambda_{0}+20A_{0}%
^{2}\rho_{20}+9A_{0}^{4}\rho_{20}^{2}}{864},~etc.
\]

In the following section we discuss our results and we draw our conclusions.

\section{Conclusions}

\label{con}

In the framework of running vacuum ($\Lambda$) models we implemented a
critical point analysis in order to study the existence and the stability of
singular solutions which describe de Sitter, radiation and matter dominated
eras. We used one of the most popular expressions for the vacuum parameter
which describes a large family of running vacuum models, namely $\Lambda
(H)=c_{0}+c_{1}H^{2}+c_{2}H^{-n}$. We showed that this functional form of
$\Lambda(H)$ allows for a number of de Sitter eras, depending on the model
parameters. The radiation and matter dominated eras are also investigated with
the method of singularity analysis. In order to recover a sequence of
radiation and matter eras, we found some interesting constraints on the cosmic fluid.

Furthermore, it is interesting to mention that a closed form solution exists
only under special conditions. Indeed, analytical solutions (see \cite{perico}
and \cite{gomez}) are possible when the cosmic fluid contains only two
components, namely one fluid (non-relativistic or radiation) and running
vacuum $\Lambda(H)$. Also, in the case of \cite{perico} we have an additional
restriction towards solving analytically the Friedamnn equations, namely we
are dealing only with even powers in $H$, hence the term $\lambda_{2}H^{-n}$
boils down to $\lambda_{2}H^{2k}$ with $n=-2k$ and $k\geq2$ [see
Eq.(\ref{fe.033})]. Despite the fact that the above special solutions are
available, we are unaware of any study concerning the critical points in
running vacuum cosmologies. Moreover, if the cosmic fluid includes radiation,
non-relativistic matter and running vacuum $\Lambda(H)$ (this situation is
close to reality) then analytical solutions are not yet available, implying
that the approximated solutions of the present study are completely new and novel.

In this context, we found families of $\Lambda(H)$ cosmologies for which new
analytical solutions are given in terms of Laurent expansions. Finally, we
showed that the Milne universe and the $R_{h}=ct$ model can be written as
perturbations deviating from a special, but unstable, $\Lambda(H)$ model.

\begin{acknowledgments}
SB acknowledges support by the Research Center for Astronomy of the Academy of
Athens in the context of the program \textquotedblleft Testing general
relativity on cosmological scales\textquotedblright\ (ref. number 200/872). AP
was financial supported by FONDECYT grant No. 3160121. AP thanks the
University of Athens for the hospitality provided while this work performed.
JDB is supported by the Science and Technology Facilities Council (STFC) of
the United Kingdom. GP is supported by the scholarship of the Hellenic
Foundation for Research and Innovation (ELIDEK grant No. 633).
\end{acknowledgments}

\end{document}